\documentclass[12pt]{iopart}

\usepackage{graphics}
\usepackage{bm}
\usepackage{dcolumn}
\usepackage{epsfig}
\usepackage{subfigure}
\usepackage{citesort}

\usepackage{fancyhdr}
\usepackage{xcolor}

\begin{document}

\title[Large deviation statistics of non-equilibrium fluctuations in a sheared model-fluid]{Large deviation statistics of non-equilibrium fluctuations in a sheared model-fluid}

\author{Pritha Dolai and Aditi Simha}

\address{Department of Physics, Indian Institute of Technology Madras, Chennai 600036, India}
\ead{\textcolor{blue}{pritha@physics.iitm.ac.in} and \textcolor{blue}{phyadt@iitm.ac.in}}
\vspace{10pt}

\begin{abstract}
We analyse the statistics of the shear stress in a one dimensional \emph{model fluid}, that exhibits a rich 
phase behaviour akin to real complex fluids under shear. 
We show that the energy flux satisfies the Gallavotti-Cohen FT across all phases in the system. 
The theorem allows us to define an effective temperature which deviates considerably from 
the equilibrium temperature as the noise in the system increases. This deviation is negligible 
when the system size is small. 
The dependence of the effective temperature on the strain rate is phase-dependent. It doesn't vary much at 
the phase boundaries. 
The effective temperature can also be determined from the large deviation function of the energy flux.
The local strain rate statistics obeys the large deviation principle and satisfies a fluctuation 
relation. It does not exhibit a distinct kink near zero strain rate because of inertia of the 
rotors in our system. 
\end{abstract}

\tableofcontents
\maketitle

\section{Introduction}
The first quantitative statement of heat production in finite systems was provided by 
the Fluctuation theorem (FT) of Evans, Cohen and Morris in 1993 ~\cite{Evans}. 
The theorem relates the probabilities of observing a positive value of the time averaged 
dissipative flux and a negative one of the same magnitude in a thermostatted dissipative system.
It was first demonstrated in a molecular dynamics
simulation of a sheared two-dimensional fluid of hard disks ~\cite{Evans}.

Several classifications of the FT exist. The  \emph{steady state} and \emph{transient} FTs differ 
in the ensemble of trajectory segments considered. In the stationary state 
Fluctuation theorem (SSFT), such as in the original work of Evans  {\it et al} ~\cite{Evans}, 
the trajectories of fixed duration $\tau$ belong to the driven steady state and the FT 
becomes valid in the limit $\tau \rightarrow \infty$. In the transient FT (TFT) ~\cite{Evans2}, the 
segments belong to a system initially in an equilibrium state evolving into a non-equilibrium 
steady state ~\cite{Searles}. 

Another classification is based on whether the fluctuation statistics are for a \emph{global} or 
\emph{local} quantity. The FT of Evans {\it et al}
was for fluctuations in the total entropy production which is a global quantity. Gallavotti and Cohen 
~\cite{Gallavotti,Galla1} provided a rigorous 
mathematical derivation of this theorem and in addition rederived a FT for the average of local observables. 
This local FT is easier to test 
experimentally as negative fluctuations, 
too rare to be observed experimentally in global quantities, are more prevalent in local quantities. Moreover,
it does not require steady state conditions to hold globally and can be applied when it holds only 
locally. 

To put our work in context, it is essential to distinguish between the FT of Evans and Searles ~\cite{Searles} and that of
Gallavotti and Cohen ~\cite{Gallavotti,Galla1}. The former is a statement of fluctuations in the dissipative flux of thermostatted
non-equilibrium states. It provides an expression for the probability of a dissipative flux in the direction opposite
to that required by the second law of thermodynamics. It requires that the initial state or distribution satisfies 
the condition of ergodic consistency. 
The Gallavotti-Cohen FT ~\cite{Gallavotti,Galla1} is more general and applies to non-equilibrium systems driven far
from equilibrium into nonlinear chaotic regimes. It depends crucially on the Chaotic Hypothesis which
imposes certain conditions on systems to which it can be applied. Our system which is a dissipative stochastic system 
driven far from equilibrium falls under this category. We show that the 
energy flux averaged over a duration $\tau$, $W_{\tau}\,$, satisfies the 
Gallavotti-Cohen FT 
\begin{equation}
 \lim_{\tau \rightarrow \infty}\frac{1}{\tau} \ln{\frac{P(+W_{\tau})}{P(-W_{\tau})}}=\beta W_{\tau} \label{FR1} 
\end{equation}
in the steady state. $\beta = (k_B \,T_{eff})^{-1}$ defines an effective temperature. 

Both FTs have been validated in several systems.
 The Evans-Searles FT has been verified in a number of molecular dynamics simulations, electrical 
resistor circuits \cite{Garnier}, and for a colloidal particle in an optical trap ~\cite{Carberry,Wang}. 
The Gallavotti-Cohen FT has 
been satisfied in many real systems such as in turbulent flows ~\cite{Ciliberto1,Ciliberto2,Bandi}, turbulent 
Rayleigh-B\'{e}nard convection ~\cite{Shang}, vertically agitated granular gas
~\cite{Feitosa} and in
a sheared micellar gel ~\cite{Majumdar}.

Our system mimics a fluid under shear at constant strain rate and exhibits a rich phase behaviour 
similar to that seen in real complex fluids under shear~\cite{Tom}. Unlike in the isoenergetic 
simulations of a sheared two dimensional fluid ~\cite{Evans}, 
the heat and energy flux have different statistics in our system and the 
dynamics is not time reversal invariant. We show that eqn.(1) is satisfied in all phases 
of the system and study the variation of the effective temperature, $T_{eff}\,$, 
with the strain rate, strength of the stochastic force and relation to changes in phase.
We show that  $T_{eff}$ can also be obtained from the large deviation function (LDF) of the energy flux.
Finally, we analyse the statistics of fluctuations in the local strain rate. 

\section{Model and phase behaviour}

\begin{figure}[!h]
\begin{center}
{
 \includegraphics[scale=0.7,keepaspectratio=true]{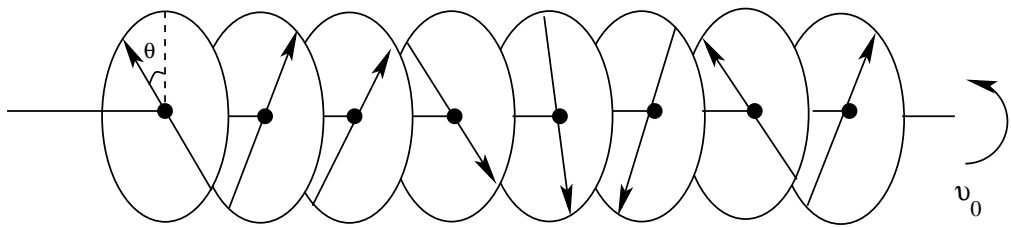}
}
\caption{The 1D driven $XY$ model. The angle $\theta$ and velocity $\dot{\theta}$ of each rotor characterizes the state
of the system at each instant.}
\label{fig:rotor}
\end{center}
\end{figure}

Our \emph{model fluid} is the classical one-dimensional $XY$ model of a lattice of spins or rotors (Figure\,~\ref{fig:rotor}), 
each of which can rotate perpendicular to the lattice. The angle $\theta_i$ of the spin 
$\bf{s_i} = (\cos \theta_i, \sin \theta_i)$ is the only degree of freedom of
each rotor or spin. Each spin interacts only with its nearest neighbours via torsional forces. In the equilibrium model these 
are forces arising from the interaction potential $U= -\sum_{1}^{N-1} {\bf{s}}_i\cdot{\bf{s}}_{i+1}$. \\ 

To generate the non-equilibrium steady states of this model, the rotors are subjected to Langevin dynamics. 
In addition to the conservative force given by $U$, frictional and stochastic forces are 
included that conserve total angular momentum. The net 
torsional force on each rotor determines its angular acceleration according to Newton's law of motion,
{\it i.e.},
\begin{equation}
I\ddot{{\theta}}_{i}=\tau_{i}=\tau_{i,i+1}+\tau_{i,i-1}\,, \label{eoms}
\end{equation}
where $I$ is the moment of inertia of each rotor which is unity in the chosen units and $\mathbf{\tau}_{i}$ the total torque acting 
on rotor $i$. This has three contributions: conservative, dissipative and random, each of which is pairwise additive and acts along $\mathbf{\hat{\theta}}_{ij}$ where 
$\theta_{ij}=\theta_{i}-\theta_{j}$ is the relative angular separation between rotors $i$ and $j$. 
\begin{equation}
\tau_i=\sum_{<ij>}\tau_{ij}^C + \tau_{ij}^D + \tau_{ij}^R 
\end{equation}
The conservative torque, 
$\tau_{i}^C=-\frac{\partial U(\theta_{ij})}{\partial \theta_{i}}$\,. The frictional torque depends only
on the relative velocity between rotors and  
has the simple form   
$\tau_{ij}^D={-}\Gamma(\dot{\theta}_{i}-\dot{\theta}_{j})$ where $\Gamma$ 
is the frictional co-efficient. The random torque $\tau_{ij}^R=\sigma \zeta_{ij}$ ,
where $\sigma$ is the amplitude of the random torque and 
$\zeta_{ij}$ is a Gaussian random variable of unit variance. 
The system is driven by rotating one of its boundaries relative to the other. This is done using Lees-Edwards 
boundary condition for the angle variable which allows us to impose a relative velocity $\upsilon_{0}$ between the 
boundaries of the system along with periodic boundary conditions so that edge effects are eliminated. 
The imposed strain rate is defined as $\dot{\gamma}=\upsilon_{0}/N$, where $N$ is the number of rotors.
The model has been used earlier to measure transition rates in non-equilibrium steady states and its 
rich phase behaviour likened to the phenomenology of real complex fluids under shear ~\cite{Baule,Mike,Tom}. A model that bears 
some similarity to ours is the sheared solid model ~\cite{Rangan}.

The equations of motion (\ref{eoms}) are integrated forward in time using a 
self-consistent Dissipative Particle Dynamics (DPD) algorithm ~\cite{Groot,Pagona}. The time step of
integration $dt$ is chosen such that the relative motion between rotors per time step is not greater than 0.24.

\begin{figure}[!h]
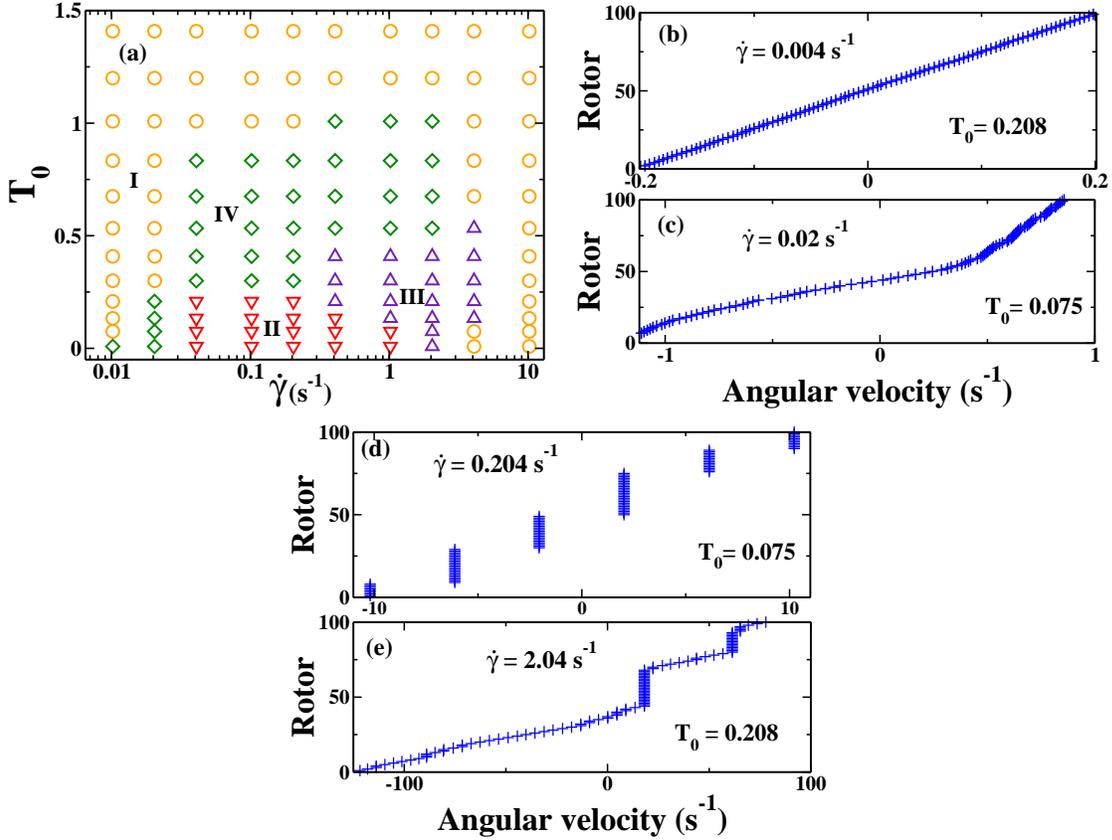

\begin{center}
 \subfigure
 {
 \includegraphics[scale=0.27,keepaspectratio=true]{fig2a.eps}
 \label{fig:phase}
}
\subfigure
{
\includegraphics[scale=0.255,keepaspectratio=true]{fig2b.eps}
\label{fig:velocity1}
}
\subfigure
{
 \includegraphics[scale=0.255,keepaspectratio=true]{fig2c.eps}
 \label{fig:velocity2}
 }
\caption{(a)Phase diagram of the model fluid for $\Gamma=0.04 s^{-1}$ and $I=1$ indicating four distinct phases. 
 (I) uniform shear phase,  (II) slip-planes, (III) solid-fluid coexistence, (IV) shear banding.
Average velocity profile in - (b)\,Phase I,\,(c)\,Phase IV\,,\,(d)\,Phase II and (e)\,Phase III.}
\end{center}
\end{figure}

The system exhibits four qualitatively distinct phases. Figure\,~\ref{fig:phase}\, depicts the various phases observed as a 
function of the imposed strain rate, $\dot{\gamma}$ and noise amplitude (expressed in terms of $T_0=\sigma^2/2 \Gamma$)
for fixed $\Gamma=0.04\,$. 
We distinguish between these phases based on the distribution of the time-averaged local strain rate 
($\langle d \theta_{i,i+1}/dt \rangle$ ) in the system. The following four phases are observed : (I) Uniform 
shear phase - the average local shear rate is 
the same through out the system. This phase exists at high $\sigma$ and $\Gamma \dot{\gamma}$ where  
thermal and frictional forces are much larger than the conservative force 
$\partial U(\theta_{ij})/\partial \theta_{i}$. Uniform shear implies a linear average velocity profile. All Newtonian fluids
shear uniformly. In complex fluids, a uniform shear flow regime always exists (Figure \,2(b)\,). (II) Slip plane phase - the shear in this phase is localized to 
a few 
neighbouring rotors (or planes) while the majority move without any relative motion between their neighbours 
as in a elastic solid (Figure \,2(d)\,).
This phase is observed when the thermal energy is small and the average torque is less than the maximum 
potential gradient. Relative motion at the slip plane produces an oscillatory torque that 
propagates in the solid region as damped torsional waves. Slip planes have been observed in 
surfactant cubic phases ~\cite{Jones}. (III) Solid-fluid
coexistence - This phase is created when the yield event at a slip plane triggers more yeild events locally  giving rise 
to a finite region thats fluid (where the rotors overcome the potential barrier), co-existing with solid regions 
(Figure \,2(e)\,). This 
co-existence is possible only when the local time averaged stress is just below the yield point. Such phases have 
been observed in foams ~\cite{Krishan,Meglio}. (IV) Shear banding - the local shear rate assumes two or more values, forming regions of 
different effective viscosities (Figure 2(c)\,). Each of these regions is called a band. Shear banding has been observed 
in polymers ~\cite{Cao,Kunita}. 
A precise characterization of these phases along with 
a mean field analysis of the model's phase behaviour can be found in  ~\cite{Tom}. 

For the range of $\dot{\gamma}$ and $\sigma$ studied (at fixed value of $\Gamma =0.04\,$), we do not see a 
turbulent phase in this system. However, we expect to see turbulence for large $\dot{\gamma}\,$, large system size, and by reducing the 
value of the frictional co-efficient $\Gamma\,$. 

We restrict our study 
here to properties of the driven steady state which is attained after 
an initial transient behaviour. In the steady state, the rate of change of the total energy  $\dot{E}=W-\dot{Q}\,$, where 
the energy flux  or power $W$ is the rate at which work is
done on the system by the driving force and the heat flux $\dot{Q}$ is the heat dissipated per unit time. Since 
the system is sheared at constant strain rate, 
$W(t)=f(t)\upsilon_{0}$ where $f$ is the net torque between neighbouring rotors on either side of the sheared boundary. The distribution of $f$ is therefore the distribution 
of the shear stress in the system. The frictional coefficient and the 
interaction potential are kept fixed in our simulations while the strain rate ($\dot{\gamma}$) and noise amplitude 
($\sigma$) are varied. 

\section{Stress fluctuations and the fluctuation theorem}
\begin{figure}[!h]
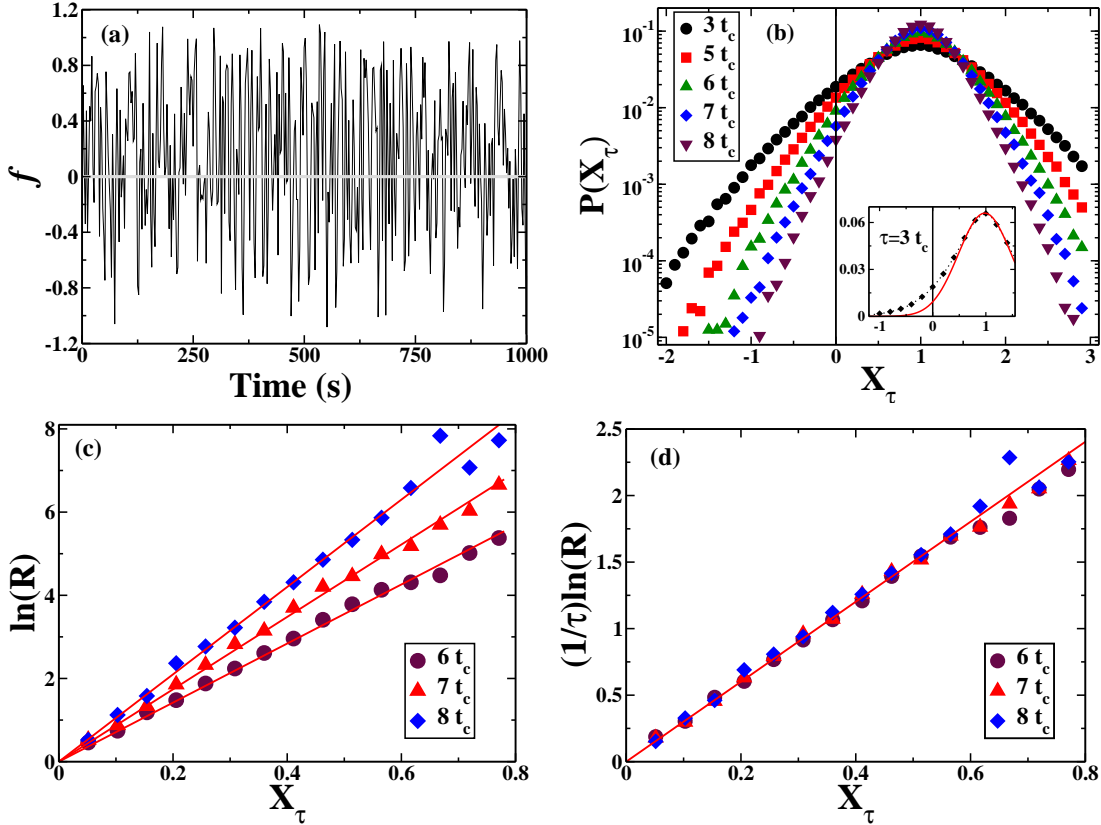

\begin{center}
 \subfigure
 {
 \includegraphics[scale=0.26,keepaspectratio=true]{fig3a.eps}
 \label{fig:fluctuation}
}
\subfigure
{
\includegraphics[scale=0.25,keepaspectratio=true]{fig3b.eps}
\label{fig:pd}
}
\subfigure
{
 \includegraphics[scale=0.26,keepaspectratio=true]{fig3c.eps}
 \label{fig:ratio}
 }
 \subfigure
{
\includegraphics[scale=0.26,keepaspectratio=true]{fig3d.eps}
\label{fig:collapse}
}
\caption{(a) Typical shear-stress fluctuations for $\dot{\gamma}$=0.327 $s^{-1}$, $\sigma=0.1$\,. 
(b) Corresponding probability distribution functions of $X_{\tau}$ for different $\tau$'s. 
(c) Plot of $\ln[P(+X_{\tau})/P(-X_{\tau})]$ vs $X_{\tau}$ for different durations $\tau$ expressed in 
terms of the correlation time $t_c\,$. Solid 
lines are the straight line fits to data. 
(d) Plot of $\frac{1}{\tau}\ln[P(+X_{\tau})/P(-X_{\tau})]$ vs $X_{\tau}$. All collapse into a straight line 
passing through the origin as shown by the fitted solid line.}
\end{center}
\end{figure}

The shear stress of the system fluctuates about a positive mean value and assumes both positive and negative values. 
Shown in Figure\,~\ref{fig:fluctuation} is a typical time evolution of the shear stress 
$f_{\tau}$ averaged over duration $\tau$ for a system of $100$ rotors in the shear banding phase. To find the probability distribution $P(f_{\tau})$, $f(t)$ is recorded as the system 
evolves in the steady state over a long duration of time (corresponding to a few hundred revolutions of each rotor) 
and over many realizations of the random torque. $2 \pi/\dot{\gamma}\,$, which is the time 
it takes for a rotor experiencing a local shear rate $\dot{\gamma}$ to complete one revolution, defines a timescale. The data 
is then averaged over different durations $\tau$\, larger than any correlation time ($t_c$) in the system. In some cases the window 
of averaging is shifted from a previous one by a 
time larger than the correlation time to improve sampling. The distribution of the resulting averaged data is $P(f_{\tau})$. 
The average energy flux into the system in a duration 
$\tau$ is $W_{\tau} = \frac{1}{\tau} \int_t^{t+\tau} W(t^{\prime})\,\mathrm{d}t^{\prime}=f_{\tau}\upsilon_{0}$; and $P(W_{\tau}/\upsilon_{0})=P(f_{\tau})$. We define 
the dimensionless quantity $X_{\tau} = W_{\tau}/ \langle W_{\tau} \rangle$. 
The distribution $P(X_{\tau})$ for different values of $\tau$ corresponding to the 
fluctuations in Fig.~\ref{fig:fluctuation}\, is shown in Fig.~\ref{fig:pd}\,. Its deviation from a Gaussian distribution is shown in the inset.   

In terms of $X_{\tau}$, eq.(\ref{FR1}) for finite $\tau$ is  
\begin{equation}
 \ln(R) \equiv \ln{\frac{P(+X_{\tau})}{P(-X_{\tau})}} =\beta \langle W_{\tau} \rangle\, X_{\tau}\tau.
\end{equation}
We set $k_B =1\,$. The straight lines in Fig.~\ref{fig:ratio} validate this and their collapse onto one 
line on scaling with $1/\tau$ (Fig.~\ref{fig:collapse}\,) validates eqn.\,(\ref{FR1}). $T_{eff}$ can be calculated 
from the slope of the collapsed line. It quantifies the probability of observing negative shear in 
the system; a large $T_{eff}$ corresponds 
to a higher probability of finding the system with negative shear stress.   

\begin{figure}[!h]
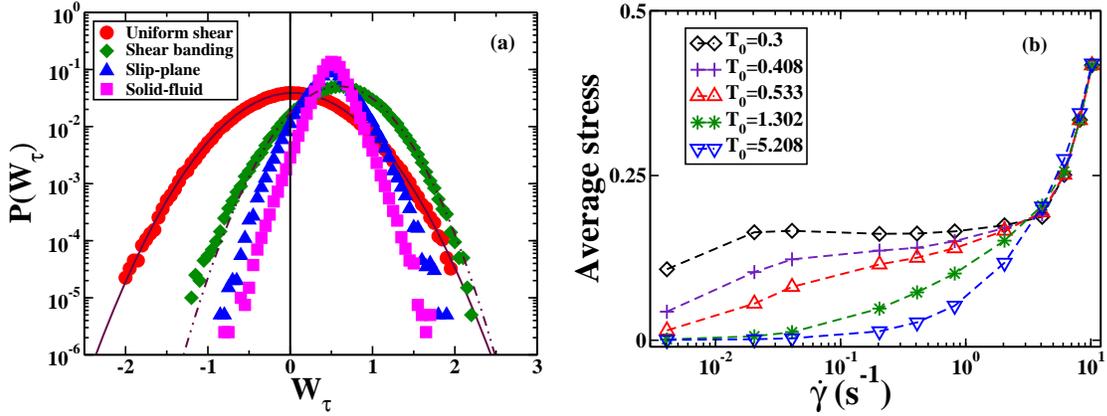

\begin{center}
 \subfigure
 {
 \includegraphics[scale=0.27,keepaspectratio=true]{fig4a.eps}
 \label{fig:pd_phase}
}
\subfigure
{
\includegraphics[scale=0.26,keepaspectratio=true]{fig4b.eps}
\label{fig:stress}
}
\caption{(a) Probability distributions in different phases. The distributions in uniform shear flow, 
shear banding, slip-plane and solid-fluid co-existence phase correspond to following sets of parameters: 
(i) $T_{0}=0.675$, $\dot{\gamma}=0.004 s^{-1}$, $t_c=4.1\,s$.  (ii) $T_{0}=0.208$, $\dot{\gamma}=0.102 s^{-1}$,
$t_c=4.4\,s$. 
(iii) $T_{0}=0.133$, $\dot{\gamma}=0.122 s^{-1}$, $t_c=3.4\,s$. 
(iv) $T_{0}=0.133$, $\dot{\gamma}=1.22 s^{-1}$, $t_c=1.7\,s$. 
(b) Average stress as a function of $\dot{\gamma}$ for different $T_{0}$\,.}
\end{center}
\end{figure}

When the fluctuations are Gaussian distributed, the probability distribution 
$P(W_{\tau}) \propto \exp {-(W_{\tau}-\langle W_{\tau} \rangle)^2/(2\sigma_{W_{\tau}}^2)}$. 
This would then imply
\begin{equation}
 \frac{P(+W_{\tau})}{P(-W_{\tau})}=\exp\,(2\langle W_{\tau} \rangle W_{\tau}/\sigma_{W_{\tau}}^2).
\end{equation}
Comparing this with the ratio of the probabilities according to the Gallavotti-Cohen FT (1), we get
\begin{eqnarray}
 2\langle W_{\tau} \rangle / \sigma_{W_{\tau}}^2= \beta\, \tau \nonumber \\
 \Rightarrow \sigma_{W_{\tau}}/\langle W_{\tau} \rangle = \sqrt{2 T/ \langle W_{\tau} \rangle \tau} \label{gauss}
\end{eqnarray}
The left hand side (LHS) is a quantity that involves the standard deviation and mean of the Gaussian
distribution. The temperature $T$ is the thermodynamic temperature when the system is in the linear response
regime. It is the effective temperature, $T_{eff}\,$, when the system is not in the linear response regime 
and the fluctuations are Gaussian distributed. The quantity on the LHS of (\ref{gauss}) is also related
to the curvature of the LDF at its minimum and its derivative at $W_{\tau}=0$ for Gaussian fluctuations.
We test if this relation holds generally when the fluctuations are not Gaussian in Sec.\,3.2\,. 

The probability distribution for $W_{\tau}$ in the various phases is shown in Fig.~\ref{fig:pd_phase}\,. The distribution at large noise amplitudes (shown by red filled circles) is 
Gaussian. The distribution in the shear banding phase is nearly Gaussian but deviates from it far away from the mean. It is non-Gaussian in the slip-plane and two phase or 
coexistence regime. 

The variation of average stress with strain rate for different $T_{0}$ is shown in Fig.~\ref{fig:stress}\,. 
For small $T_{0}$\,, the average stress plateaus before increasing sharply with 
$\dot{\gamma}$. This plateau region indicates the shear banding regime which diminishes with increasing $T_{0}$\,.
For large $T_{0}\,$, the average stress increases very slowly for small $\dot{\gamma}$ before increasing sharply 
for $\dot{\gamma}>2.0\, s^{-1}$\,. Beyond $\dot{\gamma}=4.0\, s^{-1}$ curves for different $T_{0}$ fall on each 
other in the uniform shear flow regime.

\subsection{Variation of effective temperature}
\begin{figure}[!h]
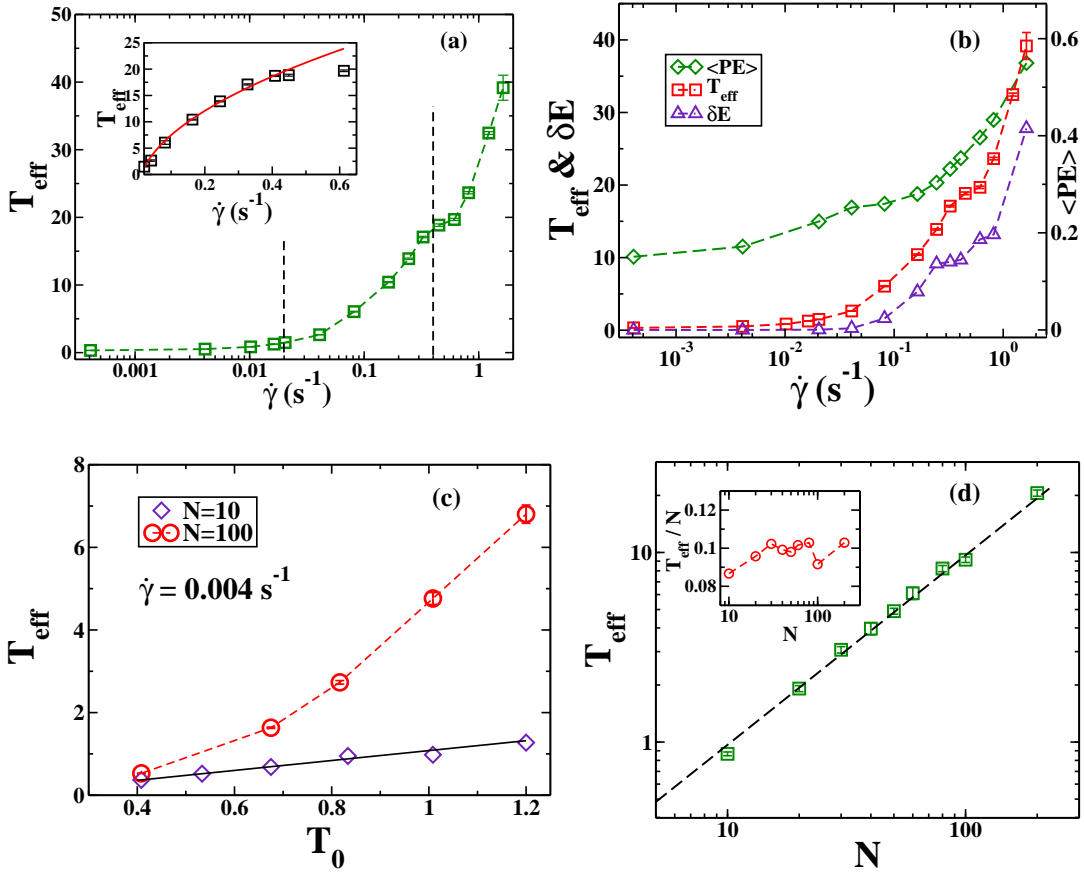

\begin{center}
 \subfigure
 {
 \includegraphics[scale=0.25,keepaspectratio=true]{fig5a.eps}
 \label{fig:teff_gdot}
}
\subfigure
{
\includegraphics[scale=0.27,keepaspectratio=true]{fig5b.eps}
\label{fig:teff_std}
}
\vskip 4mm
\subfigure
{
 \includegraphics[scale=0.275,keepaspectratio=true]{fig5c.eps}
 \label{fig:teff_sig}
 }
 \subfigure
{
 \includegraphics[scale=0.27,keepaspectratio=true]{fig5d.eps}
 \label{fig:teff_size}
 }
\caption{ (a) $T_{eff}$ as a 
function of $\dot{\gamma}$ for fixed noise amplitude $\sigma$=0.1\,. In the shear banding regime the curve is fitted 
to $\dot{\gamma}^{1/2}$ as shown in inset. The vertical lines 
indicate the boundary between different phases. 
(b) Standard deviation in energy, average potential energy and $T_{eff}$ as a
 function of $\dot{\gamma}$. 
(c) Variation of effective temperature ($T_{eff}$) with $(T_{0})$ for a fixed strain rate 
$\dot{\gamma}$=0.004 $s^{-1}$ for $N=10$ and 100. 
Black solid line indicates the corresponding equilibrium temperature $T_0$.
(d) $T_{eff}$ increases linearly with the system size. 
Inset: $T_{eff}/N$ as a function of $N$\,. }
\end{center}
\end{figure}

The variation of $T_{eff}$ with strain rate $\dot{\gamma}$ for a fixed noise amplitude, $\sigma =0.1\,$, is shown in 
Fig.~\ref{fig:teff_gdot}\,. 
In the range of $\dot{\gamma}$ covered, the system goes 
through three different phases (boundaries shown by vertical dashed lines) as can be inferred from the phase diagram; 
the uniform shear regime up to $\dot{\gamma}=0.02 \,s^{-1}$, shear banding between 
$\dot{\gamma}=0.02 - 0.4\, s^{-1}$, and the solid-fluid co-existence at higher $\dot{\gamma}$.  
Its variation with $\dot{\gamma}$ is different in each phase. The effective temperature shows a very slow increase with 
$\dot{\gamma}$ in uniform shear flow. In the shear banded phase it increases as $\sqrt{\dot{\gamma}}$ and 
much faster in the coexistence phase. At the phase boundaries $T_{eff}$ does not vary much.

The effective temperature is plotted along with the average interaction energy, $<PE>$ and the standard deviation 
of the total energy, $\delta E$, in 
Fig. ~\ref{fig:teff_std}\,. The variation of $T_{eff}$ is qualitatively similar  to 
$\delta E$. It is proportional to $\delta E$ in each of the phases albeit with a different constant of 
proportionality.  

We present our results for the variation of the effective temperature with the noise amplitude, $\sigma$, in terms of 
$T_0=\sigma^2/2\,\Gamma$, the temperature at equilibrium given 
by the Fluctuation Dissipation Theorem for the same $\sigma$. Fig.~\ref{fig:teff_sig} shows how $T_{eff}$ varies 
with $T_0$ for systems of size N = 10 and N = 100. At $N=10$\,, 
we find that $T_{eff}\approx T_{0}$ as indicated by the black solid line corresponding to $T_{eff}=T_{0}$. 
This is not true as the system size increases where it deviates considerably from $T_0$ even when the strain rate 
is small. The weak dependence for small $N$ suggests a crossover length ($L=N_0$) below which nothing happens.

An analysis of the system size dependence of $T_{eff}$, for fixed values of $\sigma$ and $\dot{\gamma}$, 
indicates that it increases linearly with system size (see Fig.\,~\ref{fig:teff_size}\,). 
We speculate that this is due to the driving speed, $\upsilon_{0}=\dot{\gamma}N\,$, which increases linearly with $N$\,. 
Plotted in the inset is $T_{eff}/N$ vs. $N$, which is nearly constant.

Except in the region corresponding to very low noise amplitudes ($\sigma<0.08$),
we have been able to show that the Gallavotti-Cohen FT is satisfied.
We speculate that at these noise amplitudes the system does not sample enough of the phase space in 
realistic time.

\subsection{Effective temperature from the large deviation function}
The effective temperature for Gaussian fluctuations can be calculated from eq.(\ref{gauss}) if the mean and standard deviation
of the distribution are known. Here we relate it to quantities in the large deviation function, its curvature 
and slope at $W_{\tau}=0$. We define the dimensionless quantity 
$X_{\tau}= W_{\tau}/\langle W_{\tau} \rangle$. $X_{\tau}$ has
the distribution,
\begin{equation}
 P(X_{\tau})= \exp[-(X_{\tau}-1)^2/2\,(\sigma_{W_{\tau}}/\langle W_{\tau} \rangle)^2],
\end{equation}
where $\langle W_{\tau} \rangle$ and $\sigma$ are the mean and standard deviation, respectively, of the 
distribution of $ W_{\tau}$. The LDF of $X_{\tau}$ is then
\begin{equation}
I(X_{\tau})=-\lim_{\tau \rightarrow \infty} \frac{1}{\tau} \ln P(X_{\tau}) = \lim_{\tau \rightarrow \infty} \frac{1}{\tau} \, \frac{(X_{\tau}-1)^2}{2\,(\sigma_{W_{\tau}}/\langle W_{\tau} \rangle)^2} 
\end{equation}
Therefore,
\begin{equation}
I^{\prime}(X_{\tau}=0) = -(\langle W_{\tau} \rangle/\sigma_{W_{\tau}})^2/\tau ,\,\, 
I^{\prime\prime}(X_{\tau}=1)=(\langle W_{\tau} \rangle/\sigma_{W_{\tau}})^2/\tau
\end{equation}

\begin{figure}[!h]
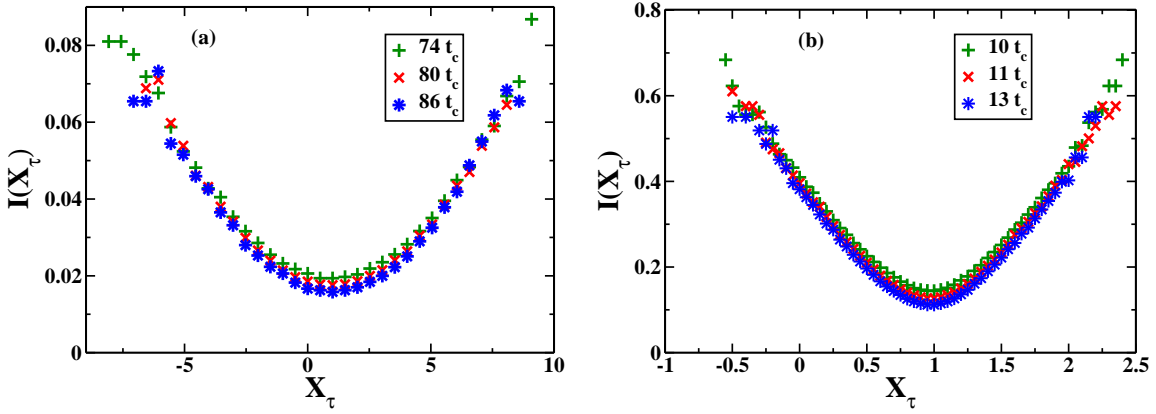

\begin{center}
 \subfigure
 {
 \includegraphics[scale=0.28,keepaspectratio=true]{fig6a.eps}
 \label{fig:ldf_gauss}
}
\subfigure
{
\includegraphics[scale=0.28,keepaspectratio=true]{fig6b.eps}
\label{fig:ldf_ng}
}
\caption{(a) LDF of $X_{\tau}$ for Gaussian fluctuations for $\sigma=0.15$ and 
$\dot{\gamma}=0.004s^{-1}$. (b) LDF for non-Gaussian fluctuations for $\sigma=0.1$ and $\dot{\gamma}=0.327s^{-1}$\,.}
\end{center}
\end{figure}

Both the curvature at the minima and the slope at $X_{\tau}=0$ of the LDF are the same, except for 
a difference in sign, and give the ratio of the mean to the standard deviation of the original 
distribution of $W_{\tau}$.  Eq.(\ref{gauss}) can be rewritten in terms of these quantities as
\begin{eqnarray}
 \frac{2\, T}{\langle W_{\tau} \rangle}= \frac{-1}{I^{\prime}(X_{\tau}=0)}\,, \label{ldf1} \\
 \frac{2\, T}{\langle W_{\tau} \rangle} = \frac{1}{I^{\prime\prime}(X_{\tau}=1)}\,. \label{ldf2}
\end{eqnarray}
Using the LDF for $X_{\tau}$, we have used eqs.~(\ref{ldf1},\ref{ldf2}) to calculate $T$ for the 
following two cases:\\
(i) $P(X_{\tau})$ is Gaussian  - $\dot{\gamma} =0.004\, s^{-1}$, $\sigma=0.15\,$, $I^{\prime}(X_{\tau}=0)=-0.002=-I^{\prime\prime}(X_{\tau}=1)$.
The yields an effective temperature $T_{eff} = 1.667$. Our earlier result for the same parameters is $1.63$.\\
(ii) $P(X_{\tau})$ is non-Gaussian -  $\dot{\gamma} = 0.327\, s^{-1}$, $\sigma=0.1\,$, $I^{\prime}(X_{\tau}=0)=-0.38\,$,
$I^{\prime \prime}(X_{\tau}=0)=-0.79\,$. Eq.(\ref{ldf1}) gives $T_{eff} = 17.0\,$. 
The effective temperature obtained from the FT is $17.09\,$.
The LDFs for the above two cases are plotted in Fig.~6\,. We find that the effective temperature can be 
obtained from the derivative of the LDF at $ W_{\tau}=0\,$.

\subsection{Statistics of the local strain rate}
The mean strain rate, $\dot{\gamma}$, imposed at the boundaries is constant in time but the local strain rate $\dot{\gamma}_{i}$ between neighbouring rotors $i$ and $i+1$ is a 
fluctuating quantity. We find the probability distribution $P(\dot{\gamma_\tau})$ of $\dot{\gamma_i}$ averaged over rotors and duration $\tau$.
\begin{figure}[!h]
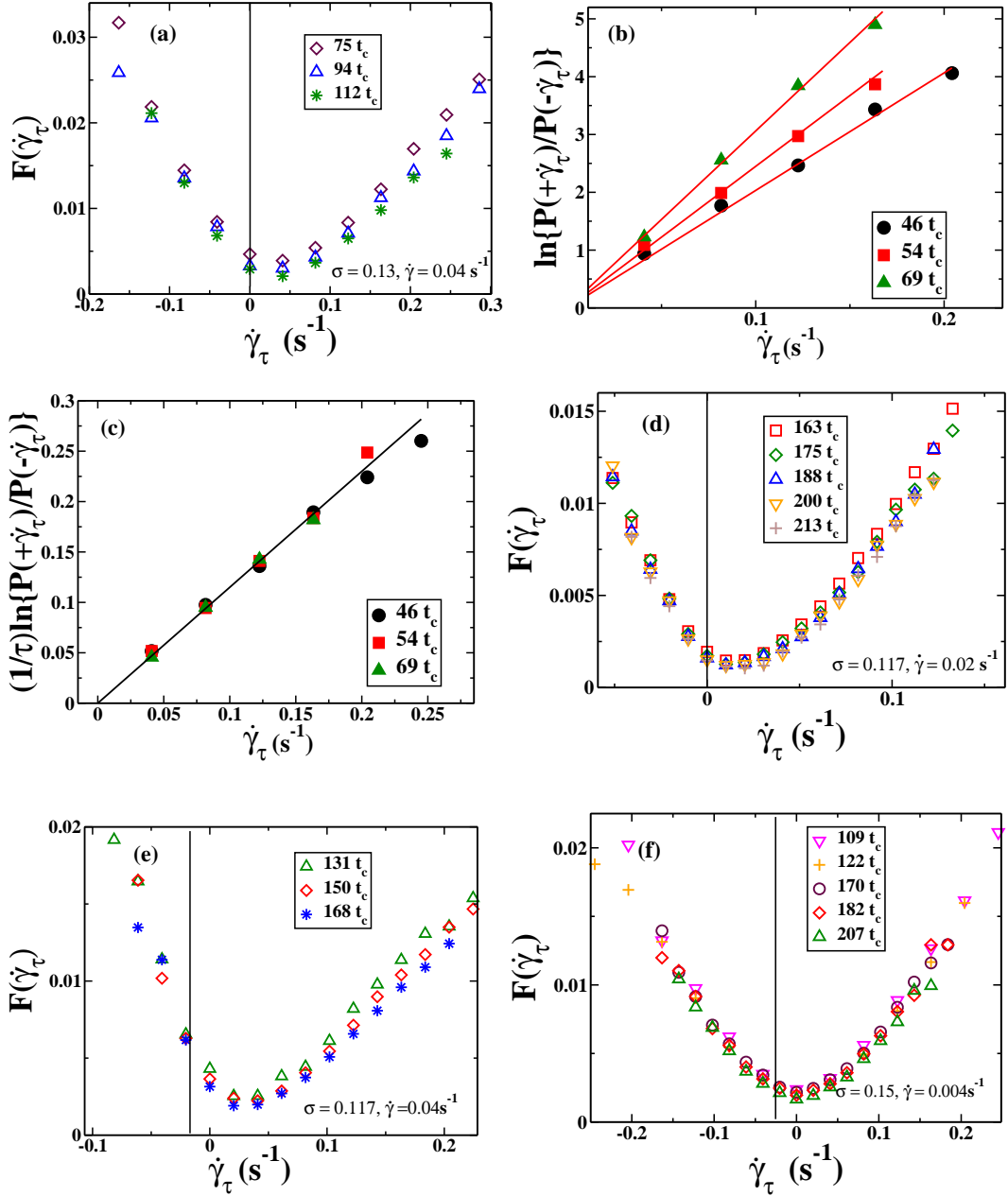

\begin{center}
 \subfigure
 {
 \includegraphics[scale=0.255,keepaspectratio=true]{fig7a.eps}
 \label{fig:ldf1a}
}
\subfigure
{
\includegraphics[scale=0.25,keepaspectratio=true]{fig7b.eps}
\label{fig:ratio1a}
}
\vspace*{0.5cm}
\subfigure
{
 \includegraphics[scale=0.25,keepaspectratio=true]{fig7c.eps}
 \label{fig:collapse1a}
 }
 \subfigure
{
 \includegraphics[scale=0.26,keepaspectratio=true]{fig7d.eps}
 \label{fig:ldf1b}
 }
 \subfigure
 {
 \includegraphics[scale=0.25,keepaspectratio=true]{fig7e.eps}
 \label{fig:ldf1c}
}
\subfigure
{
\includegraphics[scale=0.26,keepaspectratio=true]{fig7f.eps}
\label{fig:ldf1d}
}
\caption{(a) The LDF for the local strain rate for parameter values $\sigma=0.13$ and $\dot{\gamma}=0.04 s^{-1}$.
(b) Plot of $\ln[P(+\dot{\gamma}_{\tau})/P(-\dot{\gamma}_{\tau})]$ vs $\dot{\gamma}_{\tau}$ for different $\tau$ 
expressed in terms of $t_c$ for same set of parameter values. The
solid lines are the straight line fits to data. (c) All these lines of $\ln[P(+\dot{\gamma}_{\tau})/P(-\dot{\gamma}_{\tau})]$ vs $\dot{\gamma}_{\tau}$ collapse into a single straight line 
passing through the origin on scaling by $\frac{1}{\tau}$, shown by the fitted solid line. (d) The LDF for parameter values $\sigma=0.117$ and $\dot{\gamma}=0.02 s^{-1}$.
(e) The LDF for $\sigma=0.117$ and $\dot{\gamma}=0.04 s^{-1}$. (f) The large deviation function for $\sigma=0.15$ and 
$\dot{\gamma}=0.004 s^{-1}$. $t_c \approx 4\,s$ in all the plots above.}
\end{center}
\end{figure}
We define the large deviation function (LDF) for the strain rate, $F(\dot{\gamma}_{\tau})\equiv \lim_{\tau \rightarrow \infty}-(1/\tau)\,\ln{P(\dot{\gamma}_{\tau})}$ and show that 
it exists (Fig.~\ref{fig:ldf1a}\,). The antisymmetric part of the LDF (Fig.~\ref{fig:ratio1a}\,) 
obeys a fluctuation relation {\it i.e.}, $F(\dot{\gamma}_{\tau})-F(-\dot{\gamma}_{\tau}) \propto \tau \dot{\gamma}_{\tau} $ 
(Fig.~\ref{fig:collapse1a}\,). These plots are for  $\sigma=0.13\,$, $\dot{\gamma} = 0.04\,s^{-1}$. 

LDFs are plotted for 3 other sets of parameters. Fig.~\ref{fig:ldf1b} and Fig.~\ref{fig:ldf1c} show LDFs for 
$\dot{\gamma}=0.02\, s^{-1}$ and  $\dot{\gamma}=0.04\, s^{-1}$, respectively, with $\sigma$ kept constant at 0.117\,. 
Larger the strain rate $\dot{\gamma}$, more asymmetric the LDF because of the rarity of negative fluctuations. 
The LDF for $\sigma=0.15$ and $\dot{\gamma}=0.004\, s^{-1}$ is plotted in Fig.~\ref{fig:ldf1d}\,. 
This LDF is almost symmetric around zero strain rate because of the large noise amplitude. 
There is no distinctly visible kink at $\dot{\gamma}_{\tau}=0\,$ in any of the three LDFs. We speculate that this is because 
of inertia of the rotors.
A kink in the LDF has been observed earlier for entropy production in ~\cite{Speck}, and 
for the velocity of a self-propelled polar particle in ~\cite{Sriram}. In ~\cite{Speck}, the kink was attributed to 
a dynamical cross-over between a regime of high entropy production and 
a regime of low entropy production. In ~\cite{Sriram}, the kink is visible because the dynamics of particles is 
overdamped and inertia is completely ignored. 
The higher the strain rate, the greater the asymmetry of the LDF about the minimum.
A more detailed study of the LDF for the local strain rate is currently  under way. 

\section{Conclusions}
We show that the Gallavotti-Cohen FT is satisfied across all phases of the sheared \emph{model fluid} which exhibits
phases similar to real complex fluids under shear. We study the dependence of the effective temperature $T_{eff}$ 
(defined by the FT),
on the strain rate, noise amplitude and system size. $T_{eff} \approx T_0\,$, the equilibrium 
thermodynamic temperature,
at small strain rates. The linear response regime, when this happens, depends on the noise amplitude and system size.
The larger the noise amplitude, the smaller the $\dot{\gamma}$ at which the linear response regime sets in. 
$T_{eff}$ deviates considerably from 
$T_0$ as the noise amplitude increases (for fixed strain rate and system size). This deviation is negligible 
when the system size is small, suggesting that there is a crossover length (or system size) below which nothing 
happens. 
The dependence of $T_{eff}$ on $\dot{\gamma}$ is phase-dependent. It doesn't change much at the phase boundaries. 
The effective temperature can also be determined from the derivative of the LDF for the energy flux at $W_\tau=0\,$.
The local strain rate statistics obeys the large deviation principle and satisfies a fluctuation 
relation. It does not exhibit a distinct kink at zero strain rate, 
seen in other systems ~\cite{Sriram,Speck}, because of the inertia of rotors in our system.

\section*{\ack}
We thank R. M. L. Evans, Sriram Ramaswamy, Abhik Basu and S. Govindrajan for useful inputs and comments. 
We also thank P. B. Sunil Kumar for 
providing computing facilities.

\end{document}